\documentclass[runningheads]{llncs}
\usepackage{graphicx}
\usepackage{bbding}

\usepackage{algorithm}
\usepackage{algorithmicx}
\usepackage[noend]{algpseudocode}
\usepackage{colortbl}
\usepackage{subcaption}
\usepackage{booktabs}
\usepackage{url}
\usepackage{tabu}
\usepackage{multirow}
\usepackage{xspace}
\usepackage{xcolor}
\usepackage{adjustbox}
\usepackage[normalem]{ulem}
\usepackage{url}
\usepackage{amssymb}
\usepackage{amsmath}
\usepackage{enumitem}
\usepackage{hyperref}

\newcommand{\rhomin}{\ensuremath{\rho_{m}}\xspace}
\newcommand{\rhominatk}{\ensuremath{\rho_{m}^{k}}\xspace}
\newcommand{\rhoavg}{\ensuremath{\rho_{a}}\xspace}
\newcommand{\rhoatk}{\ensuremath{\rho^{k}}\xspace}
\newcommand{\rhoavgatk}{\ensuremath{\rho_{a}^{k}}\xspace}

\newcommand{\cormin}{\ensuremath{c_{m}}\xspace}
\newcommand{\dicordis}{{\sc DiCorDiS}\xspace}

\newcommand{\densthres}{\ensuremath{\delta}\xspace}
\newcommand{\corthres}{\ensuremath{\sigma}\xspace}

\newcommand{\correlationgraph}{\ensuremath{\mathcal G}\xspace}
\newcommand{\metaedges}{\ensuremath{\mathcal E}\xspace}

\newcommand{\subgraphs}{\ensuremath{\mathcal S}\xspace}
\newcommand{\summarygraph}{\ensuremath{\mathcal R}\xspace}
\newcommand{\maximalcliques}{\ensuremath{\mathcal C}\xspace}

\newcommand{\smax}{\ensuremath{s_M}\xspace}

\newcommand{\excode}{\textsc{ExCoDe}\xspace}

\newcommand{\bigO}{\mathcal{O}}

\makeatletter
 \def\@textbottom{\vskip \z@ \@plus 1pt}
 \let\@texttop\relax
\makeatother

\begin{document}

	\title{Discovering Dense Correlated Subgraphs in Dynamic Networks}
	%
	%
	\author{Giulia Preti\inst{1}{\Envelope}\and
		Polina Rozenshtein\inst{2}\thanks{work done while at Aalto University, Finland and IDS, NUS, Singapore}\and\\
		Aristides Gionis\inst{3}\and
		Yannis Velegrakis\inst{4}
	}
	
	\authorrunning{G. Preti et al.}
	\institute{ISI Foundation (\email{giulia.preti@isi.it})\and
		Amazon (\email{rozenshtein.polina@gmail.com})\and
		  KTH Royal Institute of Technology (\email{argioni@kth.se})\and
		   University of Trento and Utrecht University (\email{i.velegrakis@uu.nl})}

\maketitle
\begin{abstract}	
Given a dynamic network, where edges appear and disappear over time,
we are interested in finding sets of edges that have similar temporal behavior and form a dense subgraph. 
Formally, we define the problem as the enumeration of the maximal subgraphs that satisfy specific density and similarity thresholds.
To measure the similarity of the temporal behavior, we use the correlation between the binary time series that represent the activity of the edges. 
For the density, we study two variants based on the average degree. For these problem variants we enumerate the maximal subgraphs and compute a compact subset of subgraphs that have limited overlap. 
We propose an approximate algorithm that scales well with the size of the network, while achieving a high accuracy.
We evaluate our framework on both real and synthetic datasets. 
The results of the synthetic data demonstrate the high accuracy of the approximation and show the scalability of the framework.
\end{abstract}

\section{Introduction}
\label{sec:intro}
A popular graph-mining task is discovering dense subgraphs, i.e, portions of the graph that are densely connected. 
Finding dense subgraphs has been extensively studied in theoretical computer 
science \cite{charikar2000greedy,feige2001dense,goldberg1984finding,khuller2009finding}
and data-mining 
communities \cite{alvarez2005k,chen2012dense,gibson2005discovering,tsourakakis2013denser},
with many real-world applications \cite{angel2012dense,fratkin2006motifcut,hu2005mining}. 
For instance, dense subgraphs in a communication network indicate high-traffic regions, while in a social network indicate communities with specific interests. 
 
The majority of existing work on dense-subgraph discovery assume that the graph is static.
However, many real-life situations, e.g., social networks and road traffic networks, are highly dynamic, and hence they are better modeled by graphs that change continuously. 
In some cases, the nodes and edges undergoing structural and/or attribute changes may evolve in a convergent manner, meaning that they display a positive correlation on their behavior. 
These groups of correlated elements, especially when they involve nodes and edges that are topologically close, can represent regions of interest in the network.
In this work we are interested in discovering such patterns, i.e., in the 
\emph{discovery of correlated dense subgraphs} in dynamic networks. 
We consider graphs with edges that appear and disappear as time passes, and our goal is to identify sets of edges that show a similar behavior in terms of their presence in the graph, and at the same time, are densely connected. 
Previous works~\cite{chan2008discovering,chan2012ciforager} have considered a similar problem settings, 
but have focused on the discovery of regions of correlation valid in specific time periods, instead of throughout the whole time. 
In addition, since they create a hard partitioning of the graph, all the edges of the graph are part of the solution.
Furthermore, due to the overly restrictive model specification, 
they are heavily relying on the use of transitive metrics like the Euclidean distance.

In this work we propose a general framework for finding dense correlated subgraphs in real-world datasets, wh\-i\-ch can work with any temporal and spatial measure. Given specific density and correlation thresholds, we enumerate all maximal (meaning that the output set does not contain graphs, which are subgraphs of one another) subgraphs that satisfy the thresholds.
Furthermore, based on the observation that a dense subgraph that highly overlaps or is subsumed by a larger dense subgraph may not offer much value to the solution if the larger subgraph is already pre\-se\-nt, we produce also a more manageable and informative answer set of maximal subgraphs that are highly diverse.

Our main contributions can be summarized as follows:
\textbf{(i)} We introduce and formally define the generic problem of detecting a set of dense and correlated subgraphs in dynamic networks (Section~\ref{sec:problem}), and explain how it differs from other similar works (Section~\ref{sec:rel});
%
\textbf{(ii)} We propose two different measures to compute the density of a group of edges that change over time, which are based on the average-degree density \cite{charikar2000greedy}, and a measure to compute their correlation, based on the Pearson correlation (Section~\ref{sec:problem});
\textbf{(iii)} We develop an exact solution, called \excode, for enumerating all the subgraphs that satisfy given density and correlation thresholds. We also propose an approximate solution that scales well with the size of the network, and at the same time achieves high accuracy (Section~\ref{sec:sol});
\textbf{(iv)} As some networks naturally contain a large number of dense groups of correlated edges, we study the problem of identifying a more compact and diverse subset of results that is representative of the whole answer set. To this aim, we introduce a threshold on the maximum pairwise Jaccard similarity allowed between the edge groups in the result set, and extend our framework with an approach to extract a set of subgraphs whose pairwise overlap is less than the threshold (Section~\ref{sec:sol});
\textbf{(v)} We evaluate our framework through an extensive set of experiments on both real and synthetic datasets, confirming the correctness of the exact solution, the high accuracy of the approximate one, the scalability of the framework, and the applicability of the solution on networks of different nature (Section~\ref{sec:experiments}).
\begin{figure}[!t]
\centering
\includegraphics[width=0.7\textwidth, trim={0cm 1.5cm 0cm 1.5cm}, clip]{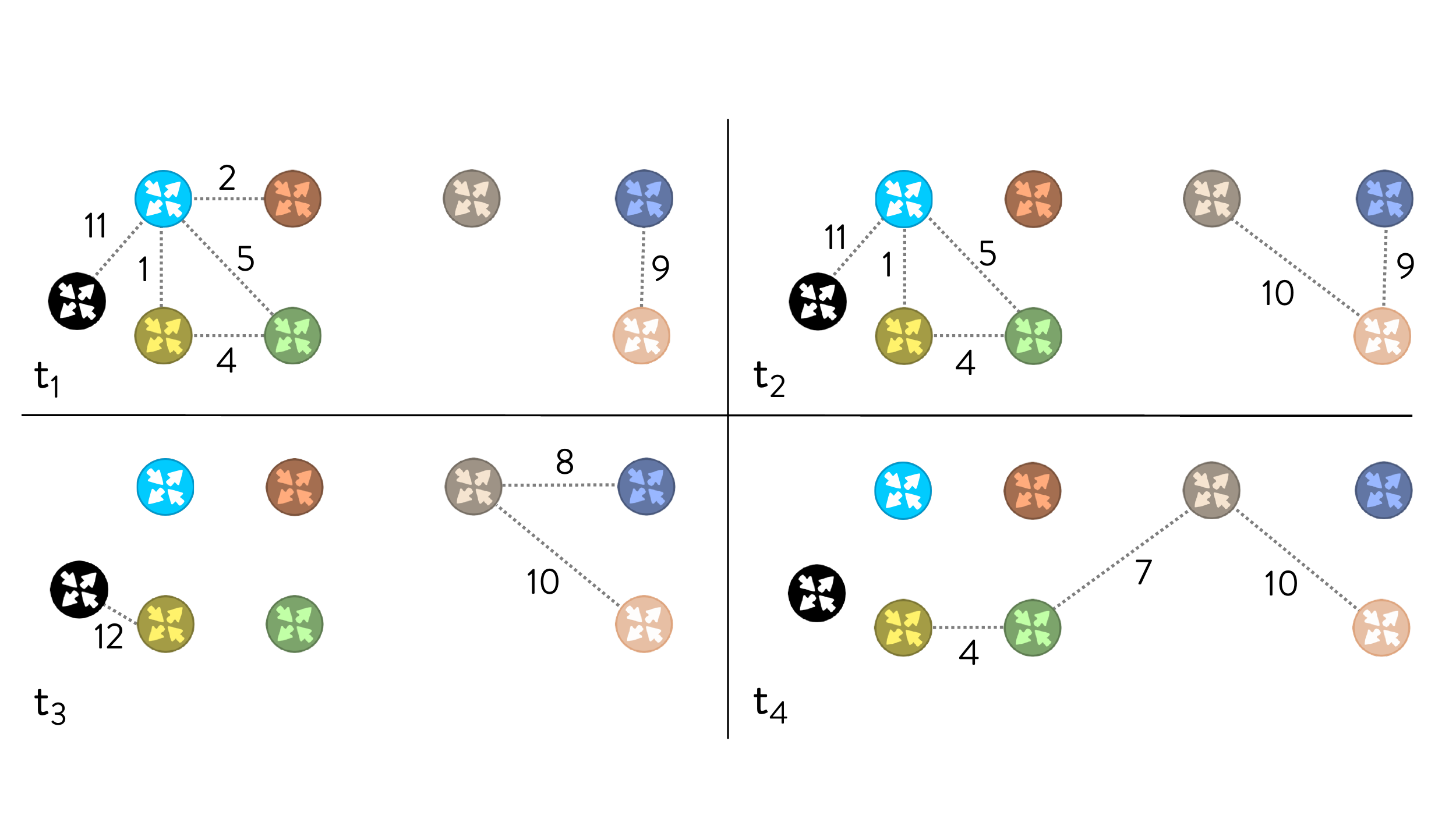}
\caption{Snapshots of a dynamic network.}
\label{fig:example}
\end{figure}

\medskip\noindent\textbf{Motivating Example.}
The Border Gateway Protocol (BGP) is the protocol used by the routers to establish how the packets are forwarded across the Internet. 
A challenge in Internet Management is to detect issues in the BGP routing topology and diagnose the human or natural disaster that caused each issue, allowing a faster recovery in the future, or preventing them from happening.
The BGP routing topology can be modeled as a graph where nodes represent routers and edges represent routing paths. As these paths can change due to reconfigurations, bottlenecks, or faults, the graph changes over time. 
Figure~\ref{fig:example} shows four snapshots of such a graph.

A fault in some router may induce changes in other portions of the graph, because all the paths traversing the faulty router must be replaced to ensure that the routing operations continue properly.
As a consequence, chan\-ge\-s in the same periods of time involving edges close in the graph are likely caused by a common cause.
For instance, the snapshots in Figure~\ref{fig:example} show that the routes $1, 5, 9$ and $11$ changed always at the same times, and hence are correlated. Among them, routes $1, 5, 11$ are close together in the network, and thus, with higher probability, they were affected by the failure of the same router. 

By focusing on the whole dense group of temporally correlated routes, a network manager is able to isolate the root causes of the faults in the topology.
However, in each snapshot of the entire BGP graph, there can be a significant number of elements experiencing a change that need to be analyzed by the manager, and in addition, not every change is associated with an anomalous event. 
Thus, there is a need for an automatic tool that can simplify the detection of the issues by finding the regions in the graph where the edges present a similar pattern of appearance, so that the analyst need to focus only on a small number of elements of the network. 
\section{Problem Statement}
\label{sec:problem}

A \emph{dynamic network} is a graph that models data that change over time.
It is represented as a sequence of static graphs ({\em snapshots} of the network).

\smallskip
\noindent
{\bf Dynamic Network.}\label{def:dynamic-network}
	Let $T \subseteq \mathcal{T}$ be a set of time instances over a domain $\mathcal{T}$. 
	A \emph{dynamic network} $D=(V,E)$ is a sequence of graphs $G_i=(V,E_i)$ with $i\in T$, 
	referred to as snapshots of the network, where
	$V$ is a set of vertices, $E_i$$\subseteq$$V$$\times$$V$ is a set of edges between vertices. 
	The set $E$ denotes the union of the edges in the snapshots, i.e., $E$=$\cup_{i{\in}T}$$E_i$.
\smallskip

We assume that all the snapshots share the same set of nodes. If a node does not interact in a snapshot, then it is present as a singleton.

Given a graph $G=(V,E)$, 
a \emph{subgraph} $H$ of $G$ is a graph $H = (V_H, E_H)$, 
such that $V_H \subseteq V$ and $E_H \subseteq E$.
In static graphs, the density of a subgraph is traditionally computed as the average degree of its nodes \cite{charikar2000greedy}:

\smallskip
\noindent
{\bf Density.}\label{def:density}
	The \emph{density} of a (static) graph $G = (V, E)$ is the average degree of its nodes, i.e., $\rho(G) = 2|E| / |V|$.
\smallskip

In the case of a dynamic network $D$, the edges of a subgraph $H$ may not exist in all the snapshots, meaning that the density may be different in each snapshot.
Therefore, we propose two approaches to aggregate the density values.
Let $G_i(H) = (V_H, E_H \cap E_i)$ denote the subgraph induced by $H$ in the snapshot $i$.
The \emph{minimum density}, denoted as $\rhomin$, is the minimum density of any subgraph induced by $H$ across the snapshots of $D$; while 
the \emph{average density}, denoted as $\rhoavg$, is the average density among these induced subgraphs.
In particular,
\vspace*{-5pt}
\begin{equation}\label{eq:den}
    \rhomin(H) = \min_{i \in T}\,{\rho(G_i(H))},\quad
    \rhoavg(H) = \frac{1}{|T|}\sum_{i \in T}{\rho(G_i(H))}.
\vspace*{-5pt}
\end{equation}
Given a density threshold $\densthres$, a subgraph $H$ is called 
$\densthres$-dense if $\rhomin(G) \geq \densthres$ or $\rhoavg(G) \geq \densthres$, respectively. 

These intuitive definitions are too strict for those practical situations where an interesting event or anomaly exhibits itself only in a small number of snapshots of the network~\cite{akoglu2015graph}.
To account for such situations, we introduce the notion of \emph{activity} and say that a subgraph $H$ is \emph{active} at time $t$ if at least $k$ edges of $H$ exist in $t$, i.e., $|E_t \cap E_H| \geq k$. Then, we relax our density definitions and compute the minimum and average density of $H$ by aggregating only over the snapshots where $H$ is active.
Let $T_{H}^{k}$ denote the subset of snapshots $H$ is active, 
i.e., $T_{H}^{k}=\{t \mid t \in T \text{ and } |E_t \cap E_H| \geq k\}$. 
We redefine Equation~\ref{eq:den} as follows:
\vspace*{-5pt}
\begin{equation}\label{eq:kden}
{\rhominatk}(H) = \min_{i \in T_{H}^{k}}\,{\rho(G_i(H))}, \quad
{\rhoavgatk}(H) = \frac{1}{|T_{H}^{k}|}\sum_{i \in T_{H}^{k}}{\rho(G_i(H))}.
\vspace*{-5pt}
\end{equation}

If $T_{H}^{k}$ is empty, then both ${\rhominatk}(H)$ and ${\rhoavgatk}(H)$ are set to 0.
We use the notation \rhoatk to refer collectively at \rhominatk and \rhoavgatk.

We say that a subgraph is \emph{correlated} if its edges are pairwise correlated. 
We therefore represent every edge as a time series over the snapshots, and measure the correlation between two edges as the Pearson correlation between the time series. Pearson correlation is widely used to detect associations between time series~\cite{fu2011review}.
However, our framework can work with any other correlation measure.
Let $\mathbf{t}(e)$ denote the time series of the edge $e$, where 
each coordinate is set to $t_i(e) = 1$ if $e$ appears in the snapshot $i$, and thus $t_i(e) = 0$ otherwise.

\smallskip
\noindent
{\bf Edge Correlation.}\label{def:correlation}
	Let $D$ $=$ $($$V$,$E$$)$ be a dynamic network, 
	$e_1,e_2\in E$ be two edges with respective time series
	$\mathbf{t}(e_1)$ $=$ $\{$$t_1(e_1)$, $\ldots$, $t_{T}(e_1)$$\}$,
	$\mathbf{t}(e_2)$ $=$ $\{$$t_1(e_2)$, $\ldots$, $t_{T}(e_2)$$\}$, and 
	$\overline{t}(e)=\frac{1}{|T|}\sum_{i=1}^{T}t_i(e)$.
	The correlation between $e_1$ and $e_2$, 
	denoted as $c(e_1,e_2)$, is the Pearson correlation between $t(e_1)$ and $t(e_2)$, i.e., 
	\vspace*{-8pt}
	\begingroup\makeatletter\def\f@size{9}\check@mathfonts
	\begin{equation*}
	c(e_1,e_2) =
	\frac
	{\sum\limits_{i=1}^{T}(t_i(e_1) - \overline{t}(e_1))(t_i(e_2) - \overline{t}(e_2))}
	{\sqrt{\sum\limits_{i=1}^{T}(t_i(e_1) - \overline{t}(e_1))^2} \sqrt{\sum\limits_{i=1}^{T}(t_i(e_2) - \overline{t}(e_2))^2}}.
	\vspace*{-5pt}
	\end{equation*}
	\endgroup
\smallskip

Given a correlation threshold $\corthres$, the edges $e_1$ and $e_2$ are considered correlated if $c(e_1,e_2) \geq \corthres$.
We define the correlation of a subgraph $H$ as the minimum pairwise correlation between its edges, i.e., 
$\cormin(H) = \min_{e_i \neq e_j \in E_H}\,{c(e_i,e_j)},$
and say that $H$ is $\corthres$-correlated if $\cormin(H) \geq \corthres$.

Our goal is to identify all the dense and correlated subgraphs in a dynamic network. 
However, since a dense correlated subgraph may contain dense correlated substructures due to the nature of the density and correlation measures used, we restrict our attention to the {\em maximal} subgraphs. 
Thus, 
given a dynamic network $D$, a density threshold $\densthres$, and a correlation threshold $\corthres$, 
we want to find all the subgraphs $H$ that are $\densthres$-dense and $\corthres$-correlated, and are not a strict subset of another $\densthres$-dense $\corthres$-correlated subgraph.

As it is often the case with problems that enumerate a complete set of solutions that satisfy given constraints, the answer set could potentially be very large and contain solutions with a large degree of overlap. 
To counter this effect, we further focus on reporting only the {\em diverse} subgraphs, which are subgraphs that differ from one another and are representative of the whole answer set.
To measure the similarity between subgraphs, we use the Jaccard similarity between their edge sets, i.e., the Jaccard similarity between the graph $G'$=$($$V'$, $E'$$)$ and $G''$=$($$V''$, $E''$$)$, denoted as  $J(G', G'')$,  is $J(G', G'')$=$|E' \cap E''|/ |E' \cup E''|$.
Then, we require that the pairwise similarities between subgraphs in the answer set are lower than a given similarity threshold $\epsilon$.
This is in line with previous work that has aimed at finding a diverse collection of dense subgraphs~\cite{galbrun2016top}.

\medskip
\noindent
{\bf Diverse Dense Correlated Subgraphs Problem [\dicordis].}
\label{prob:diverse}
Given a dynamic network $D$, 
a density threshold $\densthres$, 
a correlation threshold $\corthres$, 
and a similarity threshold $\epsilon$,
find a collection $\subgraphs$ of maximal and diverse subgraphs such that for each $H \in \subgraphs$, $H$ is $\densthres$-dense and $\corthres$-correlated, and for each distinct $H,H'\in\subgraphs$, $J(H,H')\leq\epsilon$.
\smallskip
\section{Solution}
\label{sec:sol}

To solve \dicordis, we propose a two-step appro\-ach, called \excode 
(\textbf{\underline{Ex}}tract \textbf{\underline{Co}}rrela\-te\-d \textbf{\underline{D}}ense \textbf{\underline{E}}dges).
It first identifies maximal sets of correlated edges, and 
then extracts subsets of edges that form a dense subgraph according the density measures $\rhominatk$ or $\rhoavgatk$.
The correlation of a set of edges is computed using $c_m$.

Given the dynamic network $D=(V,E)$
we create a {\em correlation graph} $\correlationgraph=(E,\metaedges)$, 
such that the vertex set of \correlationgraph is the edge set $E$ of $D$, 
and the edges of \correlationgraph are the pairs $(e_1,e_2)\in E\times E$
that have correlation $c(e_1,e_2)\geq \corthres$.
It is easy to see that a {\em maximal clique} in the correlation graph~\correlationgraph
corresponds to a maximal set of correlated edges in $D$.

The flow of \excode is illustrated in Algorithm~\ref{alg:miner}.
Starting from the dynamic network $D=(V,E)$
the algorithm first creates the correlation graph~\correlationgraph
by adding a meta-edge between two edges of $D$ if their correlation is greater than $\corthres$.
Then it enumerates all the maximal cliques in $\correlationgraph$.
This collection of maximal cliques in $\correlationgraph$ corresponds to a 
collection \maximalcliques of maximal correlated edge sets in $D$.
Finally, \textsc{FindDiverseDenseEdges}
examines each connected component in \maximalcliques 
(by using either the density measure \rhominatk or \rhoavgatk)
to identify those constituting dense subgraphs in $D$, retaining only a subset of pairwise dissimilar subgraphs according to the similarity threshold $\epsilon$. 

\smallskip
\noindent
{\bf Creation of the Correlation Graph.}
The correlation graph \correlationgraph can be built exactly, by computing the correlation $c(e_1, e_2)$ between each pair of edges $e_1, e_2 \in E$ and retaining those pairs satisfying $c(e_1, e_2) \geq \corthres$. 
However, when $D$ is large, comparing each pair of edges is prohibitively expensive, and thus we propose an approximate solution based on \emph{min-wise hashing}~\cite{broder2000min} which is described in Algorithm~\ref{alg:corrgraph}.
Here we exploit the fact that a strong correlation between two edges
implies a high Jaccard similarity of the sets of snapshots where the edges appear. We use min-wise hashing to identify sets of candidate correlated edges. 
Specifically, we use a variant of the TAPER algorithm~\cite{zhang2006finding}, 
which repeats a min-wise hashing procedure $r$ times, 
each time using $h$ independent hash functions $h_i : T \rightarrow \mathbb{N}$. 
In each run, the algorithm computes $h$ hash values for each edge and concatenates them to create a {\em hash code} for the edge (line~\ref{line:b7}). 
For each edge $e$ and each hash function $h_i$, 
the hash value $H[e][i]$ is the minimum among the values $h_i(t)$, for timestamps $t \in T$ where $e$ exists, 
i.e., $\omega_t(e) = 1$.
For efficiency purposes, the $r\,h$ hash values are computed all together by traversing once 
the set of edges of $D$ (lines~\ref{line:b2}--\ref{line:b3}).
Edges with the same \emph{hash code} are inserted into the same bucket (line~\ref{line:b8}) 
and the Pearson correlation is calculated for each pair of edges in the same bucket (line~\ref{line:b13}). 
If the correlation is greater than the correlation threshold $\corthres$, the pair is inserted in the 
edge set of the correlation graph \correlationgraph (line~\ref{line:b14}).

The algorithm requires two parameters that specify the number of runs $r$ and the number of hash functions $h$. 
Larger $r$ means less false negatives, and larger $h$ means more effective pruning. 

\setlength{\textfloatsep}{5pt}
\begin{algorithm}[!t]
\begin{algorithmic}[1]
\Require Dynamic network $D = ( V, E )$, Density function $\rhoatk$
\Require Thresholds: Correlation $\corthres$, Density $\densthres$, Size $s_M$
\Require Thresholds: Edges-per-snapshot $k$, Similarity $\epsilon$
\Ensure Diverse dense correlated maximal subgraphs $\subgraphs$
\State $\correlationgraph \gets \Call{CreateCorrelationGraph}{G,\corthres}$\label{line:a1}
\State $\maximalcliques \gets \Call{FindMaximalCliques}{\correlationgraph}$\label{line:a3}
\State $\subgraphs \gets \Call{FindDiverseDenseEdges}{D, \maximalcliques, \rhoatk, \densthres, k, s_M, \epsilon}$\label{line:a4}
\State \Return $\subgraphs$\label{line:a5}
\end{algorithmic}
\caption{\excode}\label{alg:miner}
\end{algorithm}

\begin{algorithm}[!t]
\begin{small}
\begin{algorithmic}[1]
\Require Dynamic network $D = ( V, E )$
\Require Threshold: Correlation $\corthres$
\Ensure Correlation graph $\correlationgraph = (E,\metaedges)$
\State $\mathit{cand} \gets \emptyset$; $\metaedges \gets \emptyset$
\ForAll{$e \in E$; $i \in [0, h \, r]$}\label{line:b2}
	\State $H[e][i] \gets \mathit{min}\{h_i(t)\,|\,\omega(e, t) = 1\}$\label{line:b3}
\EndFor
\ForAll{$i \in [0, r]$}
	\State $B[i] \gets \emptyset$
	\ForAll{$e \in E$}
		\State $\mathit{code} \gets H[e][i \, h : (i + 1) \, h - 1]$\label{line:b7}
		\State $B[i][code] \gets B[i][code] \cup \{e\}$\label{line:b8}
	\EndFor
\EndFor
\ForAll{$i \in [0, r]$; $b \in B[i]$}
	\ForAll{$e_1, e_2 \in B[i][b]$}
		\State $\mathit{cand} \gets \mathit{cand} \cup \{(e_1, e_2)\}$
	\EndFor
\EndFor
\ForAll{$(e_1, e_2) \in \mathit{cand}$ \textbf{ such that } $c(e_1, e_2) \geq \corthres$}\label{line:b13}
	\State $\metaedges \gets \metaedges \cup \{(e_1, e_2)\}$\label{line:b14}
\EndFor
\State \Return $\correlationgraph \gets \Call{CreateGraph}{E,\metaedges}$
\end{algorithmic}
\end{small}
\caption{\textsc{CreateCorrelationGraph} (approximate)}\label{alg:corrgraph}
\end{algorithm} 

\smallskip
\noindent
{\bf Enumeration of the Maximal Cliques.}
After the creation of the correlation graph $\correlationgraph$, the maximal groups of correlated edges are enumerated by identifying the maximal cliques in $\correlationgraph$.
To this aim, we use our implementation of the \emph{GP} algorithm of Wang et al.~\cite{wang2017parallelizing}.
Algorithm~\ref{alg:maxcliques} recursively partitions the graph into two disjoint parts and then examines each one independently using Procedure \textsc{EnumCliques}. 
In each step, it ma\-in\-tains three sets of vertices, $\mathit{anchor}$, $\mathit{cand}$, and $\mathit{not}$. 
The set $\mathit{anchor}$, initially empty, is recursively extended 
by adding a new vertex such that, at every step, 
the vertices in $\mathit{anchor}$ are all connected in the input graph $\correlationgraph$.
The set $\mathit{cand}$, initially set to $E$, 
contains the vertices that can still be used to extend $\mathit{anchor}$, 
i.e., vertices that do not belong in $\mathit{anchor}$ and are connected to every vertex in $\mathit{anchor}$.
Finally, the set $\mathit{not}$, initially empty, contains vertices already used as extensions for $\mathit{anchor}$ in the previous steps. 

\begin{algorithm}[!t]
\begin{small}
\begin{algorithmic}[1]
\Require Graph $\correlationgraph = ( E, \metaedges )$
\Ensure Set of maximal cliques $\mathcal{C}$
\State $\mathit{anc} \gets \emptyset$; $\mathit{not} \gets \emptyset$; $\mathit{cand} \gets E$
\State $\mathcal{C} \gets \Call{EnumCliques}{\mathit{anc}, \mathit{cand}, \mathit{not}}$
\State \Return $\mathcal{C}$
\Statex

\Function{EnumCliques}{$\mathit{anc}, \mathit{cand}, \mathit{not}$}
	\State $\mathcal{C} \gets \emptyset$
	\If{$\Call{isAClique}{cand}$}\label{line:d6}
		\State \Return $\{\mathit{anc} \cup \mathit{cand}\}$\label{line:d8}
	\EndIf
	\Repeat
		\State $v \gets$ vertex with smallest degree in $\mathit{cand}$\label{line:d10}
		\State $\mathit{nxAnc} \gets \mathit{anc} \cup \{v\}$
		\State $\mathit{nxCand} \gets \mathit{cand} \cap \mathit{adj}[v]$\label{line:d12}
		\State $\mathit{nxNot} \gets \mathit{not} \cap \mathit{adj}[v]$\label{line:d13}
		\If{$\nexists u \in \mathit{nxNot}$ s.t. $\forall w \in \mathit{nxCand},\,u \in \mathit{adj}[w]$}\label{line:d14}
			\State $\mathcal{C} \gets \mathcal{C} \cup \Call{EnumCliques}{\mathit{nxAnc}, \mathit{nxCand}, \mathit{nxNot}}$
		\EndIf
		\State $\mathit{cand} \gets \mathit{cand} \setminus \{v\}$\label{line:d16}
		\State $\mathit{not} \gets \mathit{not} \cup \{v\}$\label{line:d17}
	\Until{$\Call{isAClique}{cand}$}\label{line:d18}
	\If{$\nexists u \in \mathit{not}$ s.t. $\forall w \in \mathit{cand},\,u \in \mathit{adj}[w]$}
		\State $\mathcal{C} \gets \mathcal{C} \cup \{\mathit{anc} \cup \mathit{cand}\}$\label{line:d21}
	\EndIf
	\State \Return $\mathcal{C}$
\EndFunction
\end{algorithmic}
\end{small}
\caption{\textsc{FindMaximalCliques}}\label{alg:maxcliques}
\end{algorithm}

When examining the set $\mathit{cand}$ in a graph, 
the algorithm first checks if the vertices in $\mathit{cand}$ form a clique, hence returning the maximal clique $\mathit{cand} \cup \mathit{anchor}$. Otherwise, it recursively takes the vertex with smallest degree in $\mathit{cand}$ (line~\ref{line:d10}), creates a new set $\mathit{nxCand}$ of all the vertices in $\mathit{cand}$ adjacent to $v$ (line~\ref{line:d12}), and updates $\mathit{cand}$ removing $v$ (line~\ref{line:d16}). 
Set $\mathit{not}$ is updated adding vertex $v$, as it cannot be used as a partitioning anchor anymore (line~\ref{line:d17}), and a new set \emph{nxNot} is initialized adding all the vertices in $\mathit{not}$ adjacent to $v$ (line~\ref{line:d13}).
If some vertex in \emph{nxNot} is connected to all the vertices in $\mathit{nxCand}$, the recursion stops (line~\ref{line:d14}), since the vertices $\mathit{nxCand}$ cannot generate a maximal clique.
When $\mathit{cand}$ becomes a clique, the recursion stops (line~\ref{line:d18}) and the algorithm checks if the clique is maximal before inserting it into the output set (line~\ref{line:d21}). 

\begin{algorithm}[t]
\begin{small}
\begin{algorithmic}[1]
\Require Dynamic Network $D = ( V, E )$
\Require Set of maximal cliques $\mathcal{C}$, Density function $\rhoatk$
\Require Thresholds: Density $\densthres$, Size ${\smax}$, Edges-per-snapshot $k$, Similarity $\epsilon$
\Ensure	Set of diverse dense maximal subgraphs $\subgraphs$
\State $\subgraphs \gets \emptyset$; $\mathcal{P} \gets \emptyset$
\State $\mathit{CC} \gets \Call{extractCC}{\mathcal{C}}$\label{line:e2}
\ForAll{$X \in \mathit{CC}$}
	\If{$X.size$$<$${\smax}$ \textbf{and} \Call{isMaximal}{$X$, $\subgraphs$$\cup$$\mathcal{P}$} \textbf{and} \Call{isDiverse}{$X$, $\subgraphs$}}\label{line:e4}
		\State $(\mathit{flag}, R) \gets \Call{isDense}{D, X, k, \rhoatk, \densthres}$\label{line:e5}	
		\State add $X$ to $\subgraphs$ \textbf{if} $\mathit{flag} = 1$\label{line:e7}
		\State add $R$ to $\mathcal{P}$ \textbf{if} $\mathit{flag} = 0$
	\EndIf
\EndFor
\ForAll{$X \in \mathcal{P}$}\label{line:e10}
	\State add $X$ to $\subgraphs$ \textbf{if} $\Call{isMaximal}{X, \subgraphs}$ \textbf{and} \Call{isDiverse}{X, \subgraphs}\label{line:g12}
\EndFor\label{line:e12}
\State \Return $\subgraphs$
\Statex
\Function{isDiverse}{$X, S$}
	\If{$S = \emptyset$}
		\Return $\mathbf{true}$
	\EndIf
	\ForAll{$C_j \in S$}
		\If{$| C_j \cap X | / | C_j \cup X | > \epsilon$}\label{line:g19}
			\Return $\mathbf{false}$
		\EndIf
	\EndFor
	\State \Return $\mathbf{true}$
\EndFunction
\end{algorithmic}
\end{small}
\caption{\textsc{FindDiverseDenseEdges}}\label{alg:diversesub}
\end{algorithm}

\begin{algorithm}[t!]
	\begin{small}
		\begin{algorithmic}[1]
			\Require Dynamic Network $D = ( V, E )$
			\Require A set of edges $X$, Density function $\rhoatk$ 
			\Require Thresholds: Density $\densthres$, Edges-per-snapshot $k$ 
			\Ensure	$(1, \emptyset)$ if $X$ is dense; $(0, R)$ if $X$ contains the dense subsets $R$; $(-1, \emptyset)$ o.w.
			\State $K \gets \Call{kEdgeSnapshots}{X, k}$\label{line:f2}
			\If{$\rhoatk(X, K, \densthres)$}
			\Return $(1, \emptyset)$\label{line:f3}
			\EndIf
			\If{$\Call{containsDense}{X, K, k} = \emptyset$}
			\Return $(-1, \emptyset)$
			\EndIf
			\State \Return $(0, \Call{extractDense}{X, K, k})$\label{line:f5}
			\Statex
			
			\Function{containsDense}{$X, K, k$}
			\While{$\rhoatk(X, K, \densthres / 2) = \mathbf{false}$}
			\If{$X = \emptyset$ \textbf{or} $K = \emptyset$}\label{line:f27}
			\Return $\mathbf{false}$
			\EndIf
			\State $\mathit{max} \gets \Call{getMaxDeg}{X}$
			\State $n \gets \Call{getMinDegNode}{X}$
			\If{$\mathit{max} < \densthres / 2$}
			\Return $\mathbf{false}$
			\EndIf
			\State $X \gets X \setminus \mathit{adj}(n)$\label{line:f31}
			\State $K \gets \Call{kEdgeSnapshots}{X, k}$\label{line:f32}
			\EndWhile
			\State \Return $\mathbf{true}$
			\EndFunction
			\Function{extractDense}{$X, K, k$}
			\State $R \gets \emptyset$; $\mathit{Q} \gets \{(X, K)\}$ 
			\While{$Q \neq \emptyset$}
			\State extract $(Y, K')$ from $Q$
			\If{$\rhoatk(Y, K', \densthres)$}
			\State $R \gets R \cup \{Y\}$
			\ElsIf{$K' \neq \emptyset$}\label{line:f40}
			\State $\mathit{max} \gets \Call{getMaxDeg}{Y}$
			\State $N \gets \Call{getMinDegNodes}{Y}$
			\If{$\mathit{max} < \densthres$}
			$\mathbf{continue}$
			\EndIf
			\ForAll{$n \in N$}
			\State $Y \gets Y \setminus \mathit{adj}(n)$
			\State $K' \gets \Call{kEdgeSnapshots}{Y, k}$
			\State add $(Y, K')$ to $Q$ \textbf{if} $Y \neq \emptyset$
			\EndFor
			\EndIf
			\EndWhile
			\State \Return $R$
			\EndFunction
		\end{algorithmic}
	\end{small}
	\caption{\textsc{isDense}}\label{alg:dense}
\end{algorithm}

\smallskip
\noindent
{\bf Discovery of the Dense Subgraphs.}
The goal of this step is to find connected groups of edges that form a dense subgraph, using either $\rhominatk$ or $\rhoavgatk$ as density function $\rhoatk$.
Algorithm~\ref{alg:diversesub} receives in input a set of maximal cliques $\mathcal{C}$, 
each of which represents a maximal group of correlated edges. 
Since some of the edges in a clique may not be connected in the network $D$, 
the algorithm extracts all the distinct connected components from the cliques (line~\ref{line:e2}), 
before computing the density values. 
To allow a faster discovery of the maximal groups of dense edges, the connected components are sorted in descending order of their size and processed iteratively. 
If no larger or similar dense set of the current candidate $X$ has been discovered yet (line~\ref{line:e4}), and if the size of $X$ does not exceed the threshold $s_M$, the density of $X$ is computed by Algorithm~\ref{alg:dense} (line~\ref{line:e5}). 
We recall that the similarity between two sets is the Jaccard similarity, and two sets are dissimilar if the similarity is below the threshold $\epsilon$. 
On the other hand, the threshold on the maximum size controls the size of the subgraphs in the result set $\subgraphs$, as well as the complexity of Algorithm~\ref{alg:diversesub}.

Algorithm~\ref{alg:dense} describes the steps for determining if a set of edges $X$ is dense or if at least contains some dense subset.
It uses either Procedure \textsc{isMinDense} or Procedure \textsc{isAvgDense} in Algorithm~\ref{alg:density_function} to compute the density of $X$ and thus assessing if $X$ is dense.
Procedure~\textsc{isAvgDense} computes ${\rhoavgatk}$ to determine if the average density of $X$ is above the threshold $\densthres$.
The average density of $X$ is computed as the average among the average node degrees of the subgraphs $H$ induced by $X$ in all the snapshots where at least $k$ edges of $X$ are pre\-se\-nt.

On the other hand, Procedure~\textsc{isMinDense} computes ${\rhominatk}$, which expresses the density of a group of edges as the minimum between the average degrees of the induced subgraph in all the snapshots where at least $k$ edges of the group are pre\-se\-nt, and then checks if its values is above $\densthres$.
The minimum average degree is computed by iterating over the set $K$ of snapshots at least $k$ edges of $X$ present (line~\ref{line:f22}).
Nonetheless, we can stop the iteration as soon as we find a snapshot $t$ where the subgraph $H$ induced by $X$ in $t$ has average degree below the density threshold $\densthres$ (line~\ref{line:f23}), because the minimum average degree is now guaranteed to be lower than $\densthres$.
Thanks to the optimizations described in the next paragraph, the implementation of Procedure~\textsc{isAvgDense} is more efficient than that of Procedure~\textsc{isMinDense}, and thus we call the latter only when the former returns \textbf{true}, given that the average is an upper bound on the minimum.

When the density ${\rhoavgatk}(H)$ (${\rhominatk}(H)$ respectively) of the subgraph $H$ induced by $X$ is above the threshold $\densthres$, $X$ is inserted in the result set $\subgraphs$ (Algorithm~\ref{alg:diversesub} line~\ref{line:e7}).
When the density is below the threshold $\densthres$, the set $X$ is not dense; though some subset $X' \subseteq X$ may satisfy the condition ${\rhoavgatk}(H') \geq \densthres$.
Since examining all the possible subsets of $X$ is a costly operation especially when the size of $X$ is large, Algorithm~\ref{alg:dense} uses Procedure \textsc{containsDense}, which is based on a 2-approximation algorithm for the densest subgraph problem~\cite{charikar2000greedy}, to prune the search space. 
In details, Procedure~\textsc{containsDense} iteratively removes the vertex with lowest degree from the induced subgraph $H$, until it becomes empty or its density is greater than $\densthres / 2$.
Every time a vertex is removed, its outgoing edges are removed as well (line~\ref{line:f31}), and thus the set of valid snapshots $K$ must be updated (line~\ref{line:f32}). If $K$ becomes empty, any subset of $X$ will have zero density, and thus the algorithm returns \textbf{false} (line~\ref{line:f27}).  
If the maximum value of density calculated during the execution of this algorithm is below the threshold $\densthres / 2$, it holds that $X$ cannot contain a subset $X'$ with density above $\densthres$ \cite{charikar2000greedy}, and thus \textsc{containsDense} returns \textbf{false}.
Therefore, Procedure~\textsc{extractDense}, which extracts all the dense subsets in the set $X$, is invoked (line~\ref{line:f5}) only when Procedure~\textsc{containsDense} returns \textbf{true}.

\begin{algorithm}[t]
\begin{small}
\begin{algorithmic}[1]
\Require A set of edges $X$, A set of snapshots $K$
\Require Density threshold $\densthres$
\Ensure \emph{true} if $X$ is dense; \emph{false} otherwise
\Statex
\Function{isAvgDense}{$X, K, \densthres$}
	\If{$K = \emptyset$}
		\Return $\mathbf{false}$
	\EndIf
	\State let $H$ be the subgraph induced by $X$
	\State $d \gets 1/K \sum_{t \in K}{\rho(G_t(H))}$ \label{line:f10}
	\State \Return $d \geq \densthres$
\EndFunction
\Statex
\Function{isMinDense}{$X, K, \densthres$}
	\If{$K = \emptyset$}
		\Return $\mathbf{false}$
	\EndIf
	\State let $H$ be the subgraph induced by $X$
	\ForAll{$t \in K$}\label{line:f22}
		\If{${\rho(G_t(H))} < \densthres$}\label{line:f23}
			\Return $\mathbf{false}$
		\EndIf
	\EndFor
	\State \Return $\mathbf{true}$
\EndFunction

\end{algorithmic}
\end{small}
\caption{Density Functions}\label{alg:density_function}
\end{algorithm}

When Procedure \textsc{containsDense} returns \textbf{true}, Procedure \textsc{extractDense} iteratively searches for all the dense subsets in $X$. At each iteration, a subset of edges $Y$ is extracted from the queue $Q$ and its density is checked. If $Y$ is not dense but the set of valid snapshots is not empty (line~\ref{line:f40}), a new candidate is created for each vertex $n$ with lowest degree in the subgraph induced by $Y$. These candidates are then inserted into $Q$. On the other hand, when $Y$ is dense, it is inserted into the result set $R$.
At the end of Algorithm~\ref{alg:diversesub}, the maximal subsets in the set $\mathcal{P}$, which contains the elements of all the $R$ sets computed during the search, are checked for similarity with the subsets already in $\subgraphs$. Those with Jaccard similarity below $\epsilon$ with any subsets in $\subgraphs$ are finally added to $\subgraphs$ (lines~\ref{line:e10}--\ref{line:e12}).

\noindent
{\bf Computing Average Density Efficiently.}
The average density $\rhoavg(H)$ of a subnetwork $H = (V_H, E_H)$ in a dynamic network $D$ can be computed via the \emph{summary graph} of $D$ defined as the \emph{static graph} $\summarygraph = (V,E,\sigma)$ where $V$ is the set of vertices of $D$, $E$ is the union of the edges $E_i$ of all the snapshots of $D$, and $\sigma : E \mapsto \mathbb{R}$ is a weighting function that assigns, to each edge $e \in E$, a value equal to its average appearance over all the snapshots of $D$, i.e., 
$\sigma(e) = 1/|T| \sum_{i\in T} t_i(e)$.
The following proposition ensures that $\rhoavg(H)$ is equivalent to the weighted density of $H$ in the summary graph $\summarygraph$, which is defined as $w\rho(H) = 2 \sum_{e \in E_H}{\sigma(e)} / |V_H|$.

\begin{proposition}\label{pr:summary}
Given a dynamic network $D$, its summary graph $\summarygraph$, and a subnetwork $H$, it holds that $\rhoavg(H) = w\rho(H)$.
\vspace*{-4pt}
\end{proposition}
{\bf Proof.}
$$
\vspace*{-5pt}
\rhoavg(H) = \frac{1}{|T|}\sum_{t \in T}\,{\rho(G_t(H))}
= \frac{1}{|T|}\sum_{i \in T}{\left(\frac{2 | E_H \cap E_t| }{|V_H|}\right)}$$
$$ = \frac{2}{|V_H|}\frac{1}{|T|}\sum_{i \in T}{\sum_{e \in E_H}t_i(e)} 
= \frac{2}{|V_H|}\sum_{e \in E_H}{\sigma(e)} = w\rho(H).\quad\qed$$

\vspace*{-3pt}

The weighted density of $H$ in the summary graph $\summarygraph$ can be calculated significantly faster than its average density in the dynamic network $D$, since the former is obtained by summing the appearances of the edges of $H$ defined by $\sigma$, while the latter is obtained by constructing the subgraph induced by $E_H$ in each snapshot, computing the average node degree of each induced subgraph, and taking the average among those values.
Thus, Proposition~\ref{pr:summary} allows us to improve the efficiency of our algorithm when using \rhoavg (and \rhoavgatk) density function.

\smallskip
\noindent
{\bf \excode Complexity.}
The exact construction of {\correlationgraph} takes $\bigO(|E|^2)$, as it requires the computation of all the pairwise edge correlations.
The approximate solution creates $h \cdot r$ hash values for the edges in $\bigO(h \cdot r \cdot |E|)$ and compares only the edges that share at least one hash code. Even though the worst-case time complexity is still $\bigO(|E|^2)$ (every pair of edges share some hash code), practically, the actual number of comparisons is much smaller than $|E|^2$.
The time complexity of the maximal clique enumeration is $\bigO\left( |E| \cdot \kappa(\correlationgraph)\right)$, where $\kappa(\correlationgraph)$ is the number of cliques in $\correlationgraph$.
The computation of the connected components in the maximal cliques takes $\bigO(|E| \cdot \kappa(\correlationgraph))$, as it requires a visit of the network $D$ for each clique.
In the worst case, each edge of the network belongs to a different connected component, and thus Algorithm~\ref{alg:diversesub} must iterate $|E|$ times. At each iteration, it calls Procedure~\textsc{isDense} to compute the density of the current set of edges $X$ if its size is lower than $s_M$.

Procedure~\textsc{isDense} calculates the average node degree of each subgraph induced by $X$ in all the snapshots where at least $k$ edges of $X$ are present (at most $|T|$), and thus its time is bounded by $\bigO(s_M \cdot |T|)$. 
When $X$ is not dense, the algorithm further calls Procedure~\textsc{containsDense} and Procedure~\textsc{extractDense}. The former runs in $s_M$, since it removes at least one edge from $X$ at each iteration; while the latter must process all the subsets of $X$ in the worst case ($2^{s_M}$). The complexity of Algorithm~\ref{alg:diversesub} is therefore $\bigO(\kappa(\correlationgraph) \cdot |E| + |E|(s_M \cdot |T| + s_M + 2^{s_M})) = \bigO(|E|(\kappa(\correlationgraph) + s_M |T| + 2^{s_M}))$, which is also the complexity of Algorithm~\ref{alg:miner}.
\section{Experimental Evaluation}
\label{sec:experiments}

We evaluate the performance of our exact and approximate solutions in terms of accuracy and execution time. 
We also integrated our solution into a tool demonstrated at ICDMW19~\cite{preti2019excode}.

The datasets considered are $3$ real networks and $6$ randomly-generated networks, the characteristics of which are shown in Tables~\ref{fig:datasets} and~\ref{fig:randomdatasets}, respectively. They report the number of vertices $|V|$, edges $|E|$, and snapshots $|\mathcal{T}|$; 
the average node degree $\mathit{d}_a(G)$; 
the average node degree per snapshot $\mathit{d}_a(G_i)$; 
and the average number of appearances of an edge in the snapshots $\mathit{c}_a(e)$.
\textsc{haggle}~\cite{konect:chaintreau07} is a human-contact network, 
\textsc{twitter}~\cite{lahoti2018joint} is a hashtag co-occurrence network created using tweets collected from 2011 to 2016, 
and 
\textsc{mobile}~\cite{DVN/KCRS61_2015} is a network modeling calls between users made available by Telecom Italia. 
The \textsc{gaussian-x-y-z} are synthetic networks generated
using the \emph{gaussian random partition graph} generator
in the Python NetworkX library\footnote{\url{https://tinyurl.com/y5sezq73}}. 
A graph is obtained by partitioning the set of $n$ nodes into $k$ groups each of size drawn from a normal distribution $\mathcal{N}(s, s/v)$, and then adding intra-cluster edges with probability $p_\mathit{in}$ and inter-cluster edges with probability $p_\mathit{out}$.

We implemented our algorithms in Java 1.8, and run the experiments on a $24$-Core ($2.40$ GHz) Intel Xeon E5-2440 
with 188Gb RAM with Linux $3.13$, limiting the amount of memory available to 150Gb. 
In addition, we implemented a Java version of \textsc{ciForager}~\cite{chan2012ciforager}, which is the approach most related to ours. For the synthetic datasets we report results based on $100$ runs.
The code used in our experiments can be found on GitHub at \url{https://github.com/lady-bluecopper/ExCoDE}.

\begin{table}[!t]
	\setlength{\tabcolsep}{8pt}
	\caption{Real datasets.}
	\centering
	\begin{tabular}{@{\extracolsep{\fill}}l c rrr c r c r c r}
		\multirow{1}{*}{Dataset}&&\multirow{1}{*}{$|V|$}&\multirow{1}{*}{$|E|$}&\multirow{1}{*}{$|\mathcal{T}|$}&&\multicolumn{1}{c}{$\mathit{d}_a(G)$} &&\multicolumn{1}{c}{$\mathit{d}_a(G_i)$} && \multicolumn{1}{c}{$\mathit{c}_a(e)$}\\
		\midrule
		\textsc{haggle} && 274 & 2{\footnotesize K} & 90 && 15.5 && 5.2 && 5.4 \\
		\textsc{twitter-s} && 767 & 2{\footnotesize K} & 2{\footnotesize K} && 6.2 && 3 && 121.1 \\
		\textsc{twitter-m} && 1.2{\footnotesize K} & 7{\footnotesize K} & 2\footnotesize{K} && 12.1 && 3.2 && 86.4 \\
		\textsc{twitter-l} && 1.3{\footnotesize K} & 10{\footnotesize K} & 2\footnotesize{K} && 15.2 && 3.3 && 68.6 \\
		\textsc{mobile-s} && 5{\footnotesize K} & 42{\footnotesize K}& 48 && 15.3 && 4.9 && 3.8 \\
		\textsc{mobile-m} && 5{\footnotesize K}& 80{\footnotesize K} & 48 && 28.6 && 6.4 && 3.6 \\
		\textsc{mobile-l} && 5{\footnotesize K} & 118{\footnotesize K} & 48 && 41.4 && 7.5 && 3.6 \\
		\bottomrule
	\end{tabular}
	\label{fig:datasets}
	\vspace*{-5pt}
\end{table}

\begin{table*}[!t]
	\setlength{\tabcolsep}{2.5pt}
	\caption{Synthetic datasets.}
	\centering
	\begin{tabular}{@{\extracolsep{\fill}} l c lll c ll c c c c c c c ccc}
		\multirow{3}{*}{Dataset}&&\multirow{3}{*}{$|V|$}&\multirow{3}{*}{$|E|$}&\multirow{3}{*}{$|\mathcal{T}|$} &&
		\multirow{3}{*}{$p_{in}$} & \multirow{3}{*}{$p_{out}$} &&
		\multicolumn{5}{c}{\emph{independent}}&&\multicolumn{3}{c}{\emph{correlated}}\\
		\cline{10-14}\cline{16-18}
		&& &  & && & && \multicolumn{1}{c}{$d_a(G)$} &&\multicolumn{1}{c}{$d_a(G_i)$} && \multicolumn{1}{c}{$a_a(e)$} &&\multicolumn{1}{c}{$d_a(G_i)$} && \multicolumn{1}{c}{$c_a(e)$}\\
		\midrule
		\textsc{gaussian-1-7-1} && 100 & 1059 & 100 && 0.7 & 0.1 && 21.1 && 10.6 && 50 && 10.32 && 48.2\\
		\textsc{gaussian-2-7-1} && 200 & 3029 & 100 && 0.7 & 0.1 && 30.2 && 15.1 && 50.1 && 15.1 && 50.1\\
		\textsc{gaussian-3-7-1} && 300 & 6070 & 100 && 0.7 & 0.1 && 40.4 && 20.2 && 50 && 20.3 && 50.3\\
		\textsc{gaussian-1-7-3} && 100 & 1825 & 100 && 0.7 & 0.3 && 36.5 && 18.2 && 50 && 18.2 && 49.9\\
		\textsc{gaussian-2-7-3} && 200 & 6828 & 100 && 0.7 & 0.3 && 68.2 && 34 && 49.9 && 33.8 && 49.6\\
		\textsc{gaussian-3-7-3} && 300 & 14723 & 100 && 0.7 & 0.3 && 98.1&& 49 && 49.9 && 48.9 && 49.8\\
		\bottomrule
	\end{tabular}
	\label{fig:randomdatasets}
	\vspace*{-5pt}
\end{table*}

\begin{table}[!t]
	\centering
	\caption{Minimum ($\mathit{F}_m$) and average ($\mathit{F}_a$) F-score, alongside running time. 
		The values in parenthesis are the worst case values. The symbol ``-'' indicates that the algorithm was not able to terminate within $2$ days.}
	\setlength{\tabcolsep}{7pt}
	\begin{tabular}{@{\extracolsep{\fill}}l c lll c lll}
		\multirow{2}{*}{Dataset} && \multicolumn{3}{c}{\textsc{ciForager}} && \multicolumn{3}{c}{\excode}\\
		\cline{3-5}\cline{7-9}
		&& $\mathit{F}_m$ & $\mathit{F}_a$ & $t(min)$ && {$\mathit{F}_m$} & {$\mathit{F}_a$} & $t(sec)$\\
		\midrule
		\textsc{gaussian-1-7-1} && 0 & (.07) .03 & (19) .2 && (.98) 1 & (.99) 1 & (1.1) .71\\
		\textsc{gaussian-2-7-1} && 0 & (.03) .01 & (365) 4 && (.00) 1 & (.95) 1 & (2.2) 1.6\\
		\textsc{gaussian-3-7-1} && 0 & (-) .007 & (-) 44 && (.98) 1 & (.99) 1 & (8.0) 4.5\\
		\textsc{gaussian-1-7-3} && 0 & (.02) .01 & (78) .9 && (.98) 1 & (.99) 1 & (1.2) 1\\
		\textsc{gaussian-2-7-3} && 0 & (-) .003 & (-) 332 && (.97) 1 & (.99) 1 & (10) 5.5\\
		\textsc{gaussian-3-7-3} && - & (-) - & (-) - && (.98) 1 & (.99) 1 & (51) 23\\
		\bottomrule
	\end{tabular}
	\label{fig:ciforager}
\end{table}

\smallskip
\noindent
{\bf Effectiveness of the Exact Solution.}
We tested the effectiveness of our exact algorithm in detecting the actual dense groups of correlated edges in the synthetic networks \textsc{gaussian-x-7-1} and \textsc{gaussian-x-7-3}.
The correlation threshold $\corthres$ is set to $0.8$; the density threshold $\densthres$ is equal to the minimum average degree among the actual dense groups, namely $2$; and the maximum size $s_M$ is set to $\infty$ to ensure we do not miss any dense group.
We measured the accuracy in terms of the Jaccard similarity between the groups of edges discovered $\subgraphs$ and the dense groups in the ground-truth $\mathcal{G}$.
First, for each group $H \in \subgraphs$, we computed the Jaccard similarity with its closest dense group in $\mathcal{G}$, and then calculated minimum and average precision as $P_a$ and $P_m$:

$P_a = \frac{1}{\left|\subgraphs\right|}\sum\limits_{H \in \subgraphs}\max\limits_{J \in \mathcal{G}}{\textsc{Jacc}(H, J)} \quad\quad P_m = \min\limits_{H \in \subgraphs}\max\limits_{J \in \mathcal{G}}{\textsc{Jacc}(H, J)}$

Then, for each dense group $J \in \mathcal{G}$, we computed the Jaccard similarity with its closest group in $\subgraphs$, and calculated minimum and average recall $R_a$ and $R_m$:

$R_a = \frac{1}{\left|\mathcal{G}\right|}\sum\limits_{J \in \mathcal{G}}\max\limits_{H \in \subgraphs}{\textsc{Jacc}(H, J)} \quad\quad R_m = \min\limits_{J \in \mathcal{G}}\max\limits_{H \in \subgraphs}{\textsc{Jacc}(H, J)}$

Finally, we report the average and minimum F-score, as:\linebreak
$F_a = 2 (P_a \cdot R_a) / (P_a + R_a)$ and $F_m= 2 (P_m \cdot R_m)(P_m + R_m)$.

As shown in Table~\ref{fig:ciforager}, for each synthetic network \excode obtained both $F_a = 1$ and $F_m = 1$, meaning that the algorithm correctly identified all the dense groups despite the extra edges added between the groups in the networks.
We achieved lower scores only when using correlation thresholds $\corthres \leq 0.2$ for the smallest network, and $\corthres \leq 0.3$ for the others. In these cases it is more likely that some inter-group edges have a correlation greater than $\corthres$, and therefore the algorithm discovers sets of edges that are supersets of the actual dense groups.
Nonetheless, the $F_a$ score is always greater than~$0.94$, while the $F_m$ score is lower than $0.97$ only for network \textsc{gaussian-2-7-1}. 

In addition, Table~\ref{fig:ciforager} compares our approach with \textsc{ciForager}~\cite{chan2012ciforager}, the closest competitor. \textsc{ciForager} creates a partition of edges to optimize temporal (aggregated over temporal windows) correlation and spatial (all-pairs path) distance of each partition. 
We run the code with the default parameters: window length $w_l = 10$, window overlap size $w_i = 1$, clustering similarity threshold $0.25$, region similarity threshold $0.2$. 
The window length is the size of the window used to segment the sequence of graph snapshots into overlapping subsequences, while the overlap size indicates how much the subsequences overlap.
The clustering threshold decides which edges to be grouped together, and the region threshold determines how to merge the groups of edges found in different windows.

Since the output of \textsc{ciForager} is an edge partition, and thus contains edges that are not part of any dense group, the average and minimum precisions $P_a$ and $P_m$ are always low, and, as a consequence, the $F_a$ and $F_m$ are always lower than those of \excode.
Performance wise, \textsc{ciForager} is expensive since it computes the temporal distance for almost each pair of edges and for each window.
In contrast, our algorithm was able to terminate in less than a minute with every configuration and network tested.


\smallskip
\noindent
{\bf Effectiveness of the Approximate Solution.}
Our approximate algorithm trades accuracy for performance by approximating the set of $\corthres$-correlated edges. As described in Section~\ref{sec:sol}, the approximate set is obtained by the min-wise hashing technique with parameters $r$ and $h$. 
The parameter $r$ indicates the number of repetitions, so 
larger values of $r$ increase the quality of the result, at the cost of additional computation.
For the number of hash functions $h$, 
larger numbers generate more informative hash codes, 
meaning that the algorithm will cluster the edges into smaller groups. 
As the number of comparisons decreases, the running time decreases as well.

\begin{figure}[!t]
	\centering
	\includegraphics[width=10cm, trim={0cm 0cm 24cm 0cm}, clip]{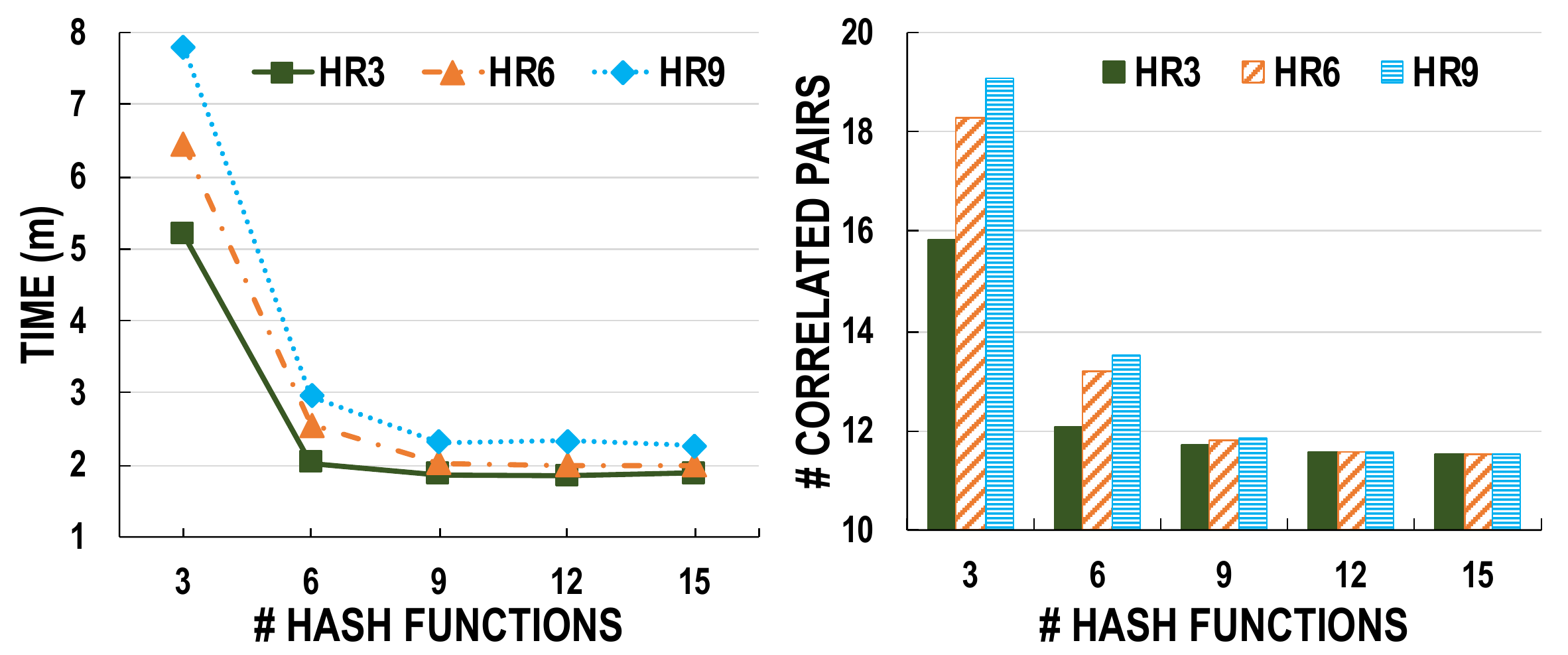}
	\caption[Tuning of the min-wise hashing parameters.]{Tuning of the min-wise hashing parameters $r$ (HR) and $h$ (HASH FUNCTIONS) running \excode \emph{approximate} in \textsc{twitter-l}.}
	\label{fig:tuning}
\end{figure}

We tested different combinations of $(r, h)$, starting from $h$$=$ $r$ $=$ $3$ and increasing their value up to $h$ $=$ $r$ $=$ $15$, and counted the number of $\corthres$-correlated edges discovered. 
Figure~\ref{fig:tuning} shows that after a period of decrease, the number of correlated pairs reached a plateau at $h = 9$, 
thus prompting us to use combinations of small values for both $r$ and $h$. 

\begin{figure}[!t]
	\centering
	\includegraphics[width=9cm, trim={0cm 0cm 24cm 0cm}, clip]{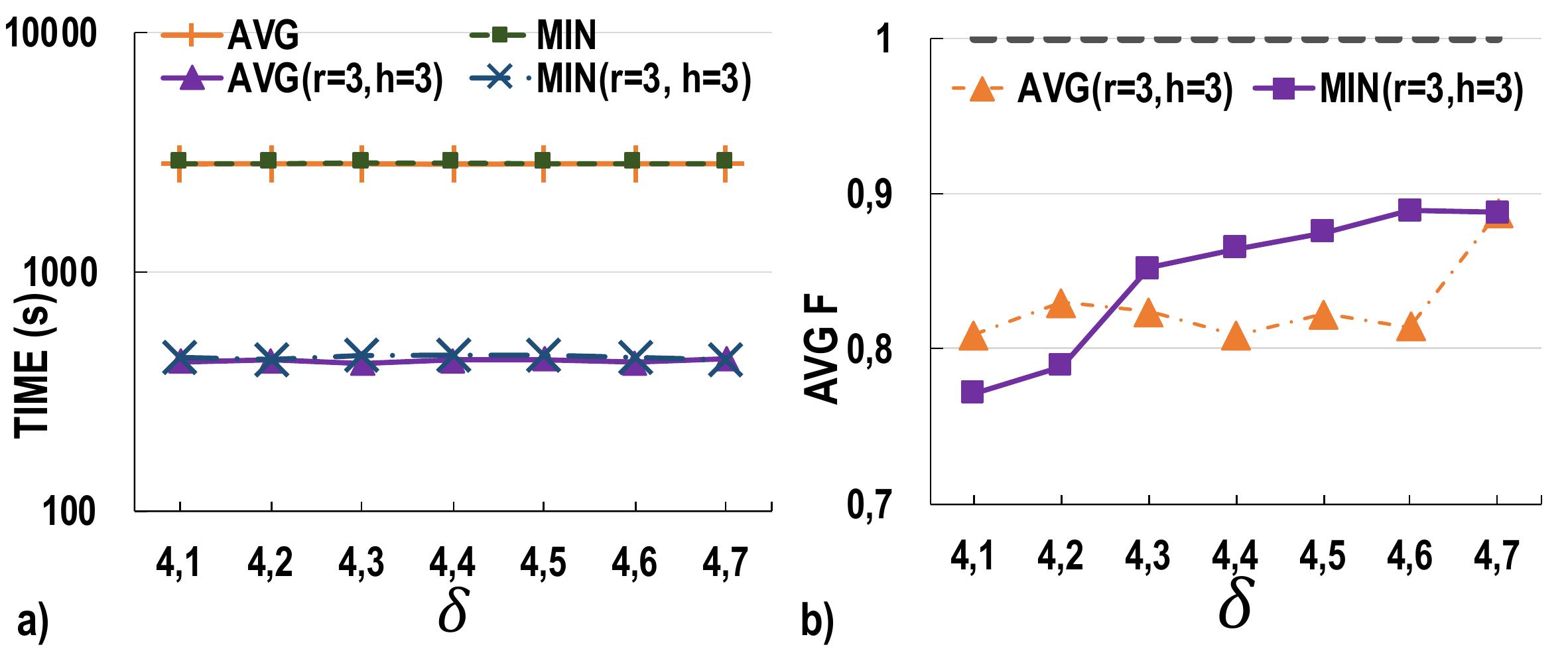}
	\vspace*{-12pt}
	\caption{Running time and average $F$ score of \excode \emph{approximate} in \textsc{mobile-m}, compared with \excode, varying $\densthres$, and using $r=3$, $h=3$, $\rhoavgatk$ (AVG) and $\rhominatk$ (MIN), and $\corthres = 0.9$.}
	\label{fig:hashing}
\end{figure}

\smallskip
\noindent
{\bf Efficiency of the Approximate Solution}.
Figure~\ref{fig:hashing} shows the performance and running time of \excode \emph{approximate} 
to find the $(0.9)$-correlated $\densthres$-dense subgraphs in the \textsc{mobile-m} network, 
using both $\rhoavgatk$ and $\rhominatk$, and varying $\densthres$.
As we can see, the approximate solution is one order of magnitude faster than the exact algorithm, 
and yet achieves a $F_a$ score of at least $0.8$ for the $\rhoavgatk$ (AVG) case, and $0.77$ for the $\rhominatk$ (MIN). 
We observed a similar behavior also in the other networks. 
As an example, in the \textsc{twitter} samples we obtained the highest $F_a$ score at high density values, while a minimum of $0.63$ at low density values.  

\begin{figure}[!t]
	\centering
	\includegraphics[width=10cm, trim={0cm 0cm 24cm 0cm}, clip]{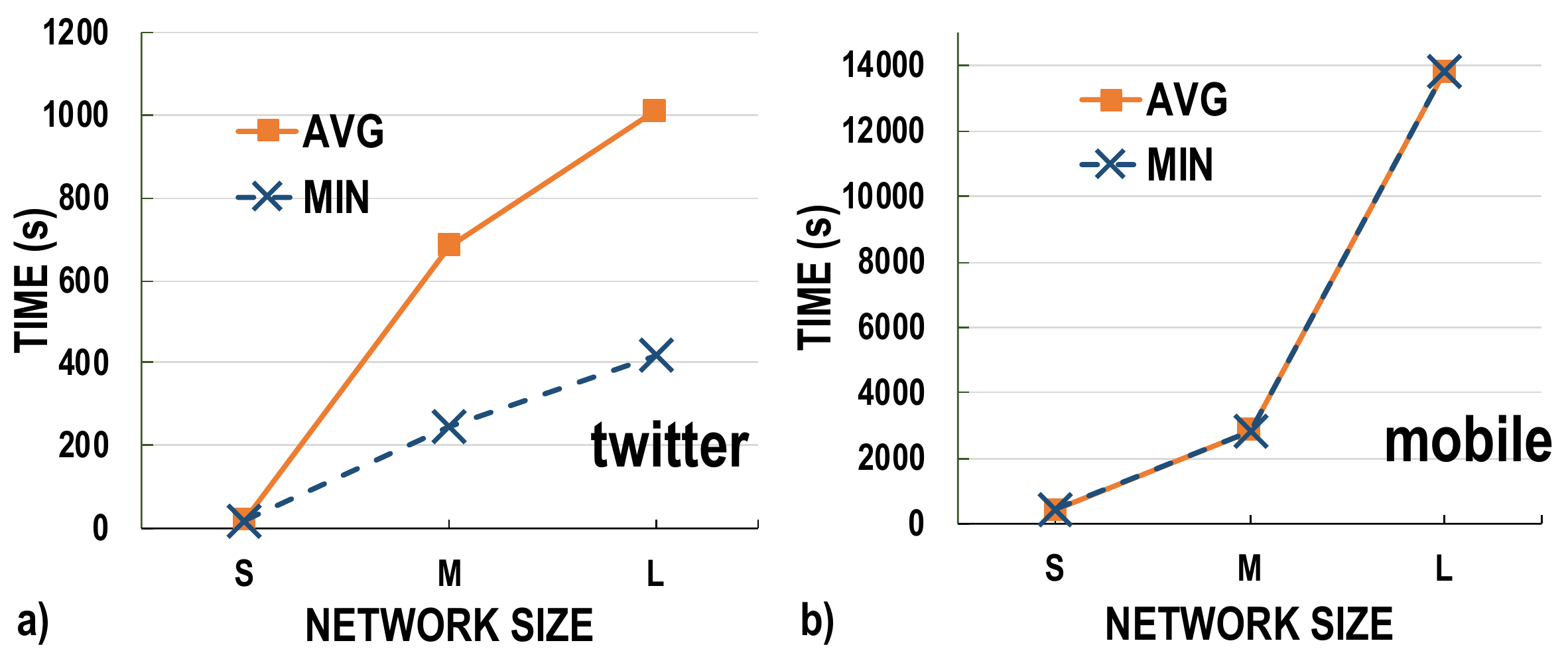}
	\caption{Scalability of \excode in \textsc{twitter} with $\corthres = 0.7$ and $\densthres = 2.6$ (a), and in \textsc{mobile} with $\corthres = 0.9$ and $\densthres=4.1$ (b), using both $\rhoavgatk$ and $\rhominatk$.}
	\label{fig:scalability}
\end{figure}

\smallskip
\noindent
{\bf Scalability}.
We tested the scalability of \excode using samples of increasing size, extracted from both the \textsc{twitter} and the \textsc{mobile} network. In the \textsc{twitter-x} samples we set $\corthres = 0.7$ and $\densthres = 2.6$, while in the \textsc{mobile-x} samples we used $\corthres = 0.9$ and $\densthres=4.1$. 
Figure~\ref{fig:scalability} shows that the running time increases exponentially for both $\rhoavgatk$ and $\rhominatk$ in the \textsc{mobile-x} samples (b), while it increases slightly slower for $\rhominatk$ in the \textsc{twitter-x} samples. 
This exponential growth is due to the \textbf{NP}-complete nature of the candidate generation task,
which requires the enumeration of the maximal cliques in the correlation graph. 
Nonetheless, these cliques can be stored and used for all the experiments where $\corthres$ is kept fixed, hence allowing a significant speed up in the performance when the user is interested in examining different combinations of the other parameters.
We also note that the different behavior of $\rhominatk$ in the \textsc{twitter-x} samples is mainly due to the sparsity of the snapshots in \textsc{twitter} and the early pruning strategy in Algorithm~\ref{alg:dense} line~\ref{line:f23} that discards a candidate group as soon as it finds a snapshot in which it is not dense.
Finally, we mention that, while \excode terminated in under $12$ minutes in the \textsc{twitter-m} sample, \textsc{ciForager} took 2.6 minutes per window (for a total of 89h using $w_l = 10$) to produce $12296$ results.

\section{Related work}
\label{sec:rel}

Our problem is close to dense subgraph mining in dynamic networks. 
Works in this field aim at retrieving the highest-scoring temporal subgraph 
\cite{ma2017fast}, 
the densest temporally compact subgraph \cite{rozenshtein2017finding}, or the group of nodes most densely connected in all the snapshots~\cite{semertzidis2018finding}.
Although they can be adapted to retrieve multiple subgraphs, the detected subgraphs are non-overlapping, and with edges that are not temporally correlated.
The enumeration of dense structures has been studied in the context of frequent subgraph mining 
\cite{abdelhamid2017incremental},
and top densest subgraph mining \cite{galbrun2016top}. When the input is a dynamic network, these groups represent subgraphs that persist over time; however, in general they are not temporally correlated.
In anomaly and fraud detection, other measures have been considered, together with the density, with the goal of finding interesting regions in the snapshots of a 
dynamic network~\cite{akoglu2015graph}.
All such works focus on the statistically significant structures, while our interest is on dense groups of edges with a similar behavior over time.
A notion of correlation has been used in approaches that characterize the event dynamics by the number of articular labels in the vicinity of spacial reference nodes~\cite{guan2012measuring}, or that compute a decay factor~\cite{yu2013anomalous}. In contrast to our work, both these approaches retrieve only anomalous nodes.
The works closest to ours are CStag~\cite{chan2008discovering} and its incremental version ciForager~\cite{chan2012ciforager}, which find regions of correlated temporal change in dynamic graphs. 
However, they partition the edges into $L$ regions, meaning that each edge is a part of the output, and hence the output can be very large and contain a number of low quality graphs. 
In contrast, we enumerate only the subgraphs with large density and high pairwise edge correlation.
\section{Conclusions}
\label{sec:conc}

We studied the problem of finding maximal dense correlated subgraphs in dynamic networks.
We proposed two measures to compute the density of a subgraph that changes over time, and one to assess the temporal correlation of its edges. 
We described a framework that uses those measures to identify such subgraphs for given density and correlation thresholds.
We extended this framework to implement the retrieval of those subgraphs that have little redundant information, specified by low Jaccard similarity.
We experimentally demonstrated the limitations of the existing solutions
and provided an approximate solution that runs in an order of magnitude faster, yet achieving a good solution quality.

\section*{Acknowledgments}
Aristides Gionis is supported by three Academy of Finland projects (286211, 313927, 317085), the ERC Advanced Grant REBOUND (834862), the EC H2020 RIA project "SoBigData++" (871042), and the Wallenberg AI, Autonomous Systems and Software Program (WASP).

\bibliographystyle{splncs04}
\bibliography{bibliography}

\end{document}